\documentclass[reprint,superscriptaddress,aps,pra,amsmath,amssymb,10pt]{revtex4-1}
\usepackage{graphicx}
\usepackage{epstopdf}
\usepackage{amsmath}
\usepackage{hyperref}
\usepackage{bm}
\usepackage{siunitx}

\def\CHR{Cr$_2$O$_3$}

\begin{document}
\title{Influence of strain and chemical substitution on the magnetic anisotropy of antiferromagnetic \texorpdfstring{\CHR}{Cr2O3}: an \textit{ab-initio} study}
\author{Sai Mu}
\thanks{Present address: Materials Science and Technology Division, Oak Ridge National Laboratory, Oak Ridge, TN 37831, USA}
\affiliation{Department of Physics and Astronomy and Nebraska Center for Materials and Nanoscience, University of Nebraska-Lincoln, Lincoln, Nebraska 68588, USA}
\author{K. D. Belashchenko}
\affiliation{Department of Physics and Astronomy and Nebraska Center for Materials and Nanoscience, University of Nebraska-Lincoln, Lincoln, Nebraska 68588, USA}

\begin{abstract}
The influence of the mechanical strain and chemical substitution on the magnetic anisotropy energy (MAE) of \CHR\ is studied using first-principles calculations. Dzyaloshinskii-Moriya interaction contributes substantially to MAE by inducing spin canting when the antiferromagnetic order parameter is not aligned with the hexagonal axis. Nearly cubic crystal field results in a very small MAE in pure \CHR\ at zero strain, which is incorrectly predicted to be negative (in-plane) on account of spin canting.
The MAE is strongly modified by epitaxial strain, which tunes the crystal-field splitting of the $t_{2g}$ triplet. The contribution from magnetic dipolar interaction is very small at any strain. The effects of cation (Al, Ti, V, Co, Fe, Nb, Zr, Mo) and anion (B) substitutions on MAE are examined. Al increases MAE thanks to the local lattice deformation. In contrast, the electronic configuration of V and Nb strongly promotes easy-plane anisotropy, while other transition-metal dopants have only a moderate effect on MAE. Substitution of oxygen by boron, which has been reported to increase the N\'eel temperature, has a weak effect on MAE, whose sign depends on the charge state of B. The electric field applied along the (0001) axis has a weak second-order effect on the MAE.
\end{abstract}

\maketitle

\section{Introduction}

Cr$_2$O$_3$ is used as the active magnetoelectric material in voltage-controlled exchange bias devices which are attractive for magnetic memory and logic applications \cite{BinekDoudin,He,Belashchenko2,Ashida,Toyoki,Belashchenko1,Kosub} due to their nonvolatility and low power consumption. The magnetic anisotropy energy (MAE) is a key parameter for such applications, which affects the thermal stability of the stored information and the coercivity of the antiferromagnet, which, in turn, controls the exchange bias, switching voltage, and switching speed. It also controls the domain wall width, which is important for domain-wall-mediated memory cells \cite{Belashchenko1}.

The easy-axis MAE of pure \CHR\ is very small: $2\times10^5$ erg/cm$^3$ (6 $\mu$eV/f.u.) at low temperatures \cite{Foner0}. The tradeoff between conflicting device requirements makes it desirable to have the ability to tune MAE in both directions.
Here we explore the MAE of \CHR\ using first-principles calculations and investigate its response to epitaxial strain, chemical substitution, and applied electric field. The paper is organized as follows. Section \ref{method} describes the computational methods. In Sec. \ref{bulk}, the MAE of bulk \CHR\ is evaluated, and different contributions are analyzed. The strain dependence of MAE in pure \CHR\ is studied in Sec. \ref{strain}.
The effects of substitutional doping on both cation and anion site on MAE are explored in Sec. \ref{dope}. Conclusions are drawn in Sec. \ref{concl}.

\section{Computational details}\label{method}

\begin{figure}
\includegraphics[width=0.48\textwidth]{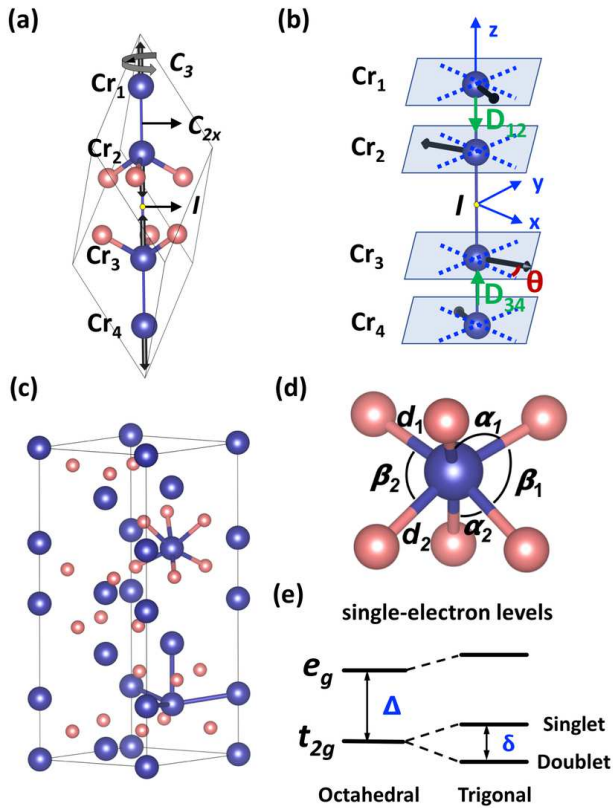}
\caption{(a) Rhombohedral primitive and (c) hexagonal unit cell of \CHR. Blue (red) spheres denote Cr (O) atoms. Here we denote the 3-fold rotation axis (rhombohedral axis or hexagonal axis) as the $z$ axis. The symmetry operations of paramagnetic \CHR\ (inversion center I, 3-fold rotation axis $C_3$, and 2-fold rotation $C_{2x}$) are labeled in panel (a). Black arrows attached to Cr atoms denote their spin orientations. (b) Spin canting arising when the primary AFM order parameter $\mathbf{L}$ is along the $x$ axis. Green arrows show the DMI vectors. (d) Local environment of the Cr atom showing bond angles and lengths. (e) Single-electron energy levels and crystal-field splitting parameters ($\Delta, \delta$). Structure visualization is done using VESTA \cite{VESTA}.}
\label{local}
\end{figure}

The dominant contribution to MAE in \CHR\ comes from the magnetocrystalline anisotropy (MCA), which is induced by spin-orbit coupling.
MCA is calculated as the difference in the total energies for two orientations of the antiferromagnetic (AFM) order parameter $\mathbf{L}$: in the (0001) plane and along the hexagonal axis (see Fig. \ref{local}).
Positive MCA corresponds to easy-axis anisotropy, i.e., $\mathbf{L}$ aligned with the hexagonal axis in the ground state.

First-principles calculations were performed using the projector augmented wave method (PAW) \cite{Blochl} implemented in the Vienna \textit{ab-initio} simulation package (VASP) \cite{Kresse,Kresse2}.
The PAW pseudopotentials for the host elements correspond to the valence-electron configuration $4s^13d^5$ for Cr and $2s^23p^4$ for O.
The exchange-correlation functional is treated in the local density approximation (LDA) \cite{Perdew0}.
The Coulomb correlations within the $3d$ shells of the transition-metal ions were described using the spherically averaged LDA+\emph{U} method \cite{Dudarev}. In this scheme, the Hamiltonian depends only on $U_\mathrm{eff}=U-J$. For Cr atoms we used $U=4.0$ eV and $J=0.6$ eV ($U-J=3.4$ eV), which give a good description of the electronic structure and magnetic properties \cite{Shi}. The plane wave cutoff energy was set to 520 eV, and a $\Gamma$-centered Monkhorst-Pack $k$-point grid \cite{Monkhorst} was used for the Brillouin zone integration. The Hellmann-Feynman forces were converged to 0.005 eV/\AA. MCA in bulk \CHR\ was computed using the rhombohedral primitive cell (see Fig.\ \ref{local}) and the tetrahedral method based on the $8\times8\times8$ $\Gamma$-centered $k$-point mesh. A demanding energy criterion, 10$^{-8}$ eV per Cr site, was employed to converge the total energy in the MCA calculations.

In order to assess the effect of epitaxial strain on the MCA of pure \CHR, we constrained the in-plane lattice parameters of the hexagonal unit cell and relaxed both the $c$ parameter and all internal degrees of freedom \cite{Mu2}. The MCA of strained \CHR\ was then calculated using the corresponding optimized rhombohedral cell.

To study the effects of alloying on MCA, we used a 30-atom hexagonal supercell (Fig.\ 1(c)) in which one Cr or O atom was substituted by another defect atom. The tetrahedron method and the $4\times4\times2$ $k$-point mesh were used for Brillouin zone integration. For the substitution of Al, we also considered a larger 90-atom unit cell, for which a $2\times2\times2$ $k$-point mesh was employed.

Substitution of one atom in a given unit cell lowers the symmetry, while the random solid solution retains the symmetry of the parent lattice.
Therefore, for each orientation of $\mathbf{L}$, it is necessary to average the energy over several supercells obtained by applying the point group symmetry operations of the original substituted site. Equivalently, one can use the same supercell but apply those symmetry operations to the \emph{spin} configuration. We use the latter procedure to evaluate MCA for all substitutions in \CHR.

The electric field control of anisotropy in ferromagnetic thin films has been widely studied \cite{Ibrahim,Ong2,Maruyama}, and the influence of electric field on the magnetic anisotropy of \CHR\ clusters embedded in MgO was also discussed \cite{Halley}. Here, we consider the effect of electric field on the MCA of bulk \CHR. The ionic displacements in the presence of electric field are evaluated using the method of Ref.\ \onlinecite{Iniguez}, and the electronic contribution to MCA is included by evaluating the electrical enthalpy \cite{Souza}.
A similar analysis has been applied in the study of the magnetoelectric effect \cite{Mu3}.

\section{Magnetocrystalline anisotropy of pure bulk \texorpdfstring{C\lowercase{r}$_2$O$_3$}{Cr2O3}} \label{bulk}

\subsection{The role of Dzyaloshinskii-Moriya interaction}
\label{DMI}

MAE corresponds to the total energy difference between spin configurations corresponding to in-plane and out-of-plane orientations of $\mathbf{L}$. For the out-of-plane orientation, the magnetic group has a $C_3$ axis passing through Cr sites, and the spin configuration is strictly collinear. However, for any other orientation, including in-plane, the Cr spins can tilt thanks to the Dzyaloshinskii-Moriya interaction (DMI) \cite{Dzyaloshinskii,Moriya}.

We use the labeling of the Cr spins in the primitive cell shown in Fig.\ \ref{local}a. According to Moriya's rule \cite{Moriya}, the DMI vector $\mathbf{D}_{12}$ for the pair of nearest-neighbor Cr atoms points along the hexagonal axis, as shown in Fig.\ \ref{local}b. For the spin pairs connected by the inversion center, such as atoms 2 and 3 in Fig.\ \ref{local}b, the DMI vector vanishes.

The primary order parameter in \CHR\ is $\mathbf{L}=\boldsymbol{\mu}_1-\boldsymbol{\mu}_2+\boldsymbol{\mu}_3-\boldsymbol{\mu}_4$, where the Cr sites are labeled as shown in Fig.\ \ref{local}a. The $R\bar3c$ space group allows the free energy invariant $L_{x}L_{y}'-L_{y}L_{x}'$, where $\mathbf{L}'=\boldsymbol{\mu}_1+\boldsymbol{\mu}_2-\boldsymbol{\mu}_3-\boldsymbol{\mu}_4$ \cite{Borovik}. Microscopically, this invariant is generated by DMI and has a contribution from the nearest-neighbor Cr pairs. Clearly, if the primary $\mathbf{L}$ has a finite in-plane component, an orthogonal component of $\mathbf{L}'$ is induced as a secondary order parameter, resulting in the canting of the spins away from collinearity, as shown in Fig.\ \ref{local}b. This effect is similar to weak ferromagnetism generated by another invariant $L_{x}^{\prime\prime}M_y-L_{y}^{\prime\prime}M_x$, where $\mathbf{L}^{\prime\prime}=\boldsymbol{\mu}_1-\boldsymbol{\mu}_2-\boldsymbol{\mu}_3+\boldsymbol{\mu}_4$ is the primary order parameter in hematite \cite{Borovik}. Note that the relations $\boldsymbol{\mu}_1=-\boldsymbol{\mu}_4$ and $\boldsymbol{\mu}_2=-\boldsymbol{\mu}_3$ remain valid for an arbitrary orientation of $\mathbf{L}$.

The presence of spin canting according to the $\mathbf{L}'$ pattern is confirmed by explicit calculations allowing for spin non-collinearity when $\mathbf{L}$ lies in the (0001) plane. We found the induced transverse spin moments on Cr sites of about 0.09 $\mu_B$, which amounts to a 1.7$^{\circ}$ canting angle.

Spin canting lowers the energy of the in-plane spin configuration and thereby reduces the MCA. To emphasize the role of this effect, in the following we discuss, in addition to the MCA energy $K$ obtained with full (noncollinear) spin relaxation, the corresponding energy difference calculated while keeping the spin configuration collinear, denoted as $K_{col}$.

The experimental value of MCA in \CHR\ is about 6 $\mu$eV/f.u.\ \cite{Foner0}. For pure, unstrained \CHR, with the theoretically optimized structure, we found $K_{col}=2.5$ $\mu$eV/f.u. and $K=-44\ \mu$eV/f.u. A similar value of $K=-52$ $\mu$eV/f.u.\ was obtained in the experimental structure \cite{Artman}.
Thus, the full calculation allowing for spin canting erroneously predicts easy-plane anisotropy. It is reasonable to attribute this error to the failure of the DFT$+U$ calculation to give the correct deviation of the crystal-field splitting on Cr atoms from cubic. As we show below, a small change of about 30 meV in the crystal-field splitting is enough to make MCA positive. Despite this overall offset in MCA, we expect the trends in its variation under strain or substitution to be reasonably captured.

\subsection{Analysis of the spin-orbit coupling energy}
\label{kso}

To gain insight into the origin of MCA, especially in the presence of an impurity, it is useful to consider the spin-orbit coupling energy for each atom, $E_\mathrm{SO}(\mathbf{n})=\sum_{\sigma\sigma'}\mathop\mathrm{Tr}\xi_{\sigma\sigma'} \hat{\mathbf{L}}\cdot\hat{\mathbf{S}}_{\sigma\sigma'}\hat\rho_{\sigma'\sigma}(\mathbf{n})$, along with the individual terms in the summation over spin indices. Here the trace is over the orbital indices, $\xi_{\sigma\sigma'}$ is the $l$-diagonal spin-orbit coupling parameter, $\hat\rho(\mathbf{n})$ is the density matrix projected onto a local basis for a specific atom, calculated for $\mathbf{L}\parallel\mathbf{n}$, in the reference frame where the axis $z$ is parallel to $\mathbf{n}$. Within second-order perturbation theory, $K\approx K_\mathrm{SO}=[E_\mathrm{SO}(\perp)-E_\mathrm{SO}(\parallel)]/2$ \cite{Ke1}, where the summation over all atoms is implied.

Table \ref{decompose} lists the four contributions to the spin-orbit energy on one Cr site in \CHR, for two orientations of $\mathbf{L}$.
We see that the diagonal majority-spin and the off-diagonal contributions to $E_\mathrm{SO}$ are comparable. On the other hand, the off-diagonal contribution dominates in the MCA as a result of spin canting.
Table \ref{decompose} also shows the that the small orbital moment on Cr comes largely from the majority-spin states.

\begin{table}[htb]
\caption {Spin decomposition of the spin orbit energy $E_\mathrm{SO}(\mathbf{n})$ ($\mu$eV), the orbital magnetic moment $M_l(\mathbf{n})$ (10$^{-3}$ $\mu_B$) on Cr, and their anisotropies. $\mathbf{n}$ is the orientation of the AFM order parameter.
}
\begin{tabular}{|l|cccc|c|}
\hline
 &$\uparrow\uparrow$ & $\downarrow\downarrow$ & $\downarrow\uparrow$ & $\uparrow\downarrow$& Total            \\
\hline
$E_\mathrm{SO}(\mathbf{z})$            & $-1012.91 $ & $-85.75$& $-1311.70$  & $-1311.70$  & $-3722.07$    \\
$E_\mathrm{SO}(\mathbf{x})$            & $-1011.24$ & $-87.13$& $-1334.62$  & $-1334.62$  &  $-3767.61$   \\
$K_\mathrm{SO}$   &  $ 0.84 $&  $-0.69$     & $-11.46$       & $-11.46$ & $-22.77$    \\
\hline
$M_l(\mathbf{z})$            & 41.80 & $-4.80$ &   &   & 37.00    \\
$M_l(\mathbf{x})$            & 41.93 & $-4.83$ &   &   & 37.11    \\
$\Delta M_l$        & -0.13 & 0.03      &         &     & -0.11       \\
\hline
\end{tabular}
\label{decompose}
\end{table}

\section{\texorpdfstring{C\lowercase{r}$_2$O$_3$}{Cr2O3} under epitaxial strain} \label{strain}

\begin{figure} [htb]
\includegraphics[width=0.54\textwidth]{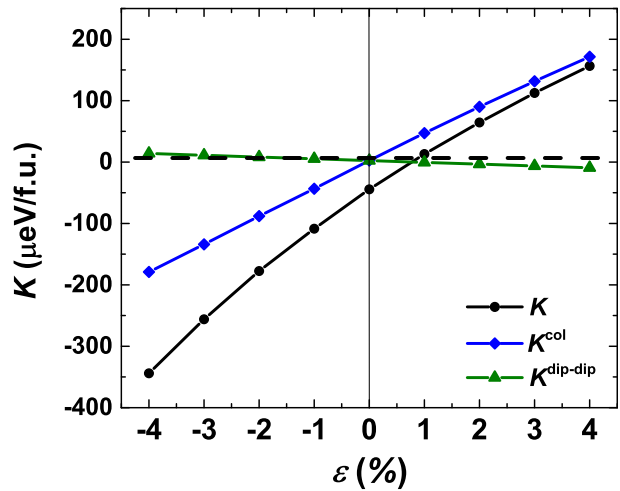}
\caption{Dependence of $K$ (black circles) and $K_{col}$ (blue diamonds) on the in-plane strain under biaxial stress in \CHR. Green triangles: dipole-dipole contribution to MAE. Dashed line: experimental MAE of unstrained \CHR. }
\label{epi}
\end{figure}

Given that the crystal field in \CHR\ is close to cubic, the MCA should be strongly affected by mechanical strain. Here, we consider biaxial stress ($\sigma_{xx}=\sigma_{yy}=\sigma$, $\sigma_{zz}=0$), which appears in a thin film that is laterally constrained by an epitaxial substrate. This is implemented by fixing the in-plane lattice parameter $a$ and relaxing both $c$ and the internal atomic positions. Figure \ref{epi} shows the dependence of MCA on the magnitude of epitaxial strain. In the linear region, $K$ and $K_{col}$ increase under tensile epitaxial strain at a rate of 61 and 44 $\mu$eV/f.u. per 1\% of strain, respectively.

The energy change due to spin canting (which accounts for the difference between $K_{col}$ and $K$) can be written as $\Delta E(\theta)=-D\theta+\frac12 J\theta^2$, assuming small canting angle $\theta$, where $D$ and $J$ are the effective DMI and exchange coupling parameters corresponding to the canting mode. The small contribution from the magnetic anisotropy is included in $J$. Minimization over $\theta$ gives the equilibrium values $\theta_0=D/J$ and $\Delta E_0=-\frac{1}{2}D^2/J$, where $\Delta E_0=K-K^{col}$.

\begin{figure} [htb]
\includegraphics[width=0.48\textwidth]{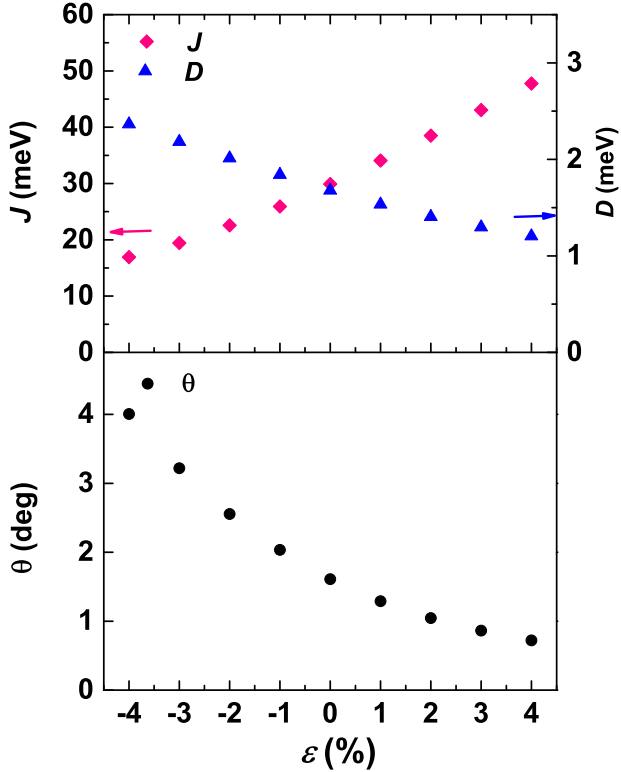}
\caption{(a) Effective exchange ($J$) and DMI parameters ($D$), and (b) the spin canting angle $\theta$ as a function of epitaxial strain.}
\label{canting_strain}
\end{figure}

Figure \ref{canting_strain} shows the parameters $J$, $D$, as well as the canting angle $\theta$, as a function of epitaxial strain. Tensile strain ($\varepsilon>0$) increases $J$ while decreasing $D$, which results in the reduction of $\theta$, which also decreases the difference between $K$ and $K^{col}$, as seen in Fig. \ref{epi}.

\begin{figure}
\includegraphics[width=0.5\textwidth]{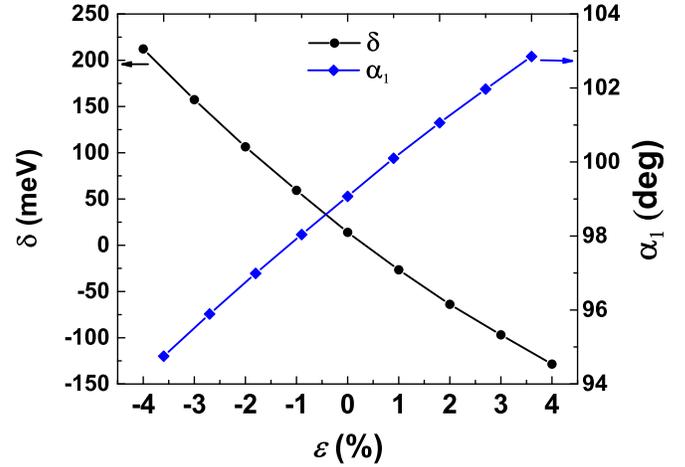}
\caption{Epitaxial strain-dependent $t_{2g}$ level splitting $\delta$ and the O-Cr-O bonding angle $\alpha_1$.}
\label{t2g}
\end{figure}

The increase of MCA under tensile strain is due to its effect on the crystal-field splitting $\delta$ of the $t_{2g}$ states (Fig. \ref{local}d). This parameter is controlled by the O-Cr-O bond angles. There are two inequivalent angles, denoted $\alpha_1$ and $\alpha_2$ in Fig. \ref{local}d, with both O atoms belonging to the same crystallographic layer, and two ($\beta_1$ and $\beta_2$) with O atoms in different layers. Tensile strain increases the $\alpha_i$ angles while decreasing $\beta_i$, as seen in Table \ref{bonds}, which also lists the bond lengths. Figure \ref{t2g} also shows that the splitting of the $t_{2g}$ states at the $\Gamma$ point, which represents the crystal-field parameter $\delta$, decreases under increasing strain and passes through zero at a small positive value of $\varepsilon$.

\begin{table*}[htb]
\caption {Cr-O bond length $d_i$ (\AA) in \CHR\ under epitaxial strain $\varepsilon$ or with Al substitution $x$. $\alpha_{1,2}$, $\beta_{1,2}$: O-$X$-O bond angles (deg) shown in Fig.\ \ref{local}. Lattice parameters (\AA) and the c/a ratio are also given. }
\begin{tabular}{|c|c|c|c|c|c|c|c|c|c|c|c|c|c|c|c|c|c|}
\hline
$\varepsilon$ (\%) & $ -4\%$ & $ -3\%$& $-2\%$ & $ -1\%$& $0\%$ & $ +1\%$& $+2\%$& $ +3\%$&$+4\%$& \multicolumn{2}{c|}{$x=2.8$\%} &\multicolumn{2}{c|}{x=0.083}  \\
\hline
Ligand  & Cr$-$O  & Cr$-$O & Cr$-$O & Cr$-$O&Cr$-$O  & Cr$-$O & Cr$-$O & Cr$-$O& Cr$-$O &Cr$^{NN}-$O & Cr$^{NNN}-$O  &Cr$^{NN}-$O  &Cr$^{NNN}-$O    \\
\hline
$d_1$            & 1.929	&1.937 &	1.944	&1.952	&1.960	&1.969	&1.979	&1.989 &	1.999& 1.961 & 1.957  & 1.958 & 1.951      \\
$d_2$            & 1.991	&1.995&	1.998	&2.002	&2.010&	2.010	&2.014	&2.018	&2.022  & 1.998 & 2.008  & 1.995 & 2.015   \\
\hline
$\alpha_1$     & 94.75	&95.89	&96.99	&98.04	&99.07	&100.10	&101.06	&101.97	&102.85& 98.61 & 101.12 & 99.23 & 102.07   \\
$\alpha_2$     & 79.76	&80.11	&80.48	&80.85	&81.23	&81.63	&82.02	&82.41	&82.81  & 79.07 & 81.17  & 79.15 & 82.48    \\
$\beta_1$      & 90.54	&89.59	&88.67	&87.77	&86.87	&85.96	&85.09	&84.25	&83.42  & 87.78 & 86.41  & 87.39 & 82.86     \\
$\beta_2$      & 94.12	&93.45	&92.78	&92.13	&91.47	&90.83	&90.21	&89.60	&89.01  &93.02  & 91.12  & 92.60  & 90.97     \\
\hline
$a$         & 4.7470	& 4.7964	& 4.8459	&4.8953	&4.9440	&4.9942	&5.0437	&5.0931 &	5.1426  &   \multicolumn{2}{c|} {4.2753}  &    \multicolumn{2}{c|} {4.9260}     \\
$c/a$         &  2.9776	 & 2.9166	& 2.8570	&2.7992	&2.8570	&2.6851	&2.6307	&2.5781	&2.5267  &   \multicolumn{2}{c|} {2.7401}  &    \multicolumn{2}{c|} {2.7413}   \\
\hline
\end{tabular}
\label{bonds}
\end{table*}

The collinear approximation $K_{col}$ exhibits the same trend as a function of strain as the full calculation $K$. To understand the origin of this trend, consider a localized independent-electron model of the Cr $3d$ shell.
The Cr site symmetry $C_3$ splits the $t_{2g}$ triplet into a doublet and a singlet separated by $\delta$, which is typically less than 100 meV, while the energy of the $e_g$ doublet $\Delta$ is of order 1 eV. As a basis set, we use linear combinations of cubic harmonics resulting from this splitting. In this basis, the crystal field Hamiltonian for the majority-spin $3d$ electrons is
\[ H_{cf} = \left( \begin{array}{ccccc}
0 & 0 & 0 & 0 & 0\\
0 & 0 & 0 & 0 & 0 \\
0 & 0 & \delta &0 & 0    \\
0 & 0 & 0 &\Delta & 0    \\
0 & 0  & 0 & 0 & \Delta
 \end{array} \right).\]
The full model Hamiltonian is obtained by adding the Hund exchange splitting $H_J=\epsilon \hat{\mathbf{S}}\mathbf{n}$ and spin-orbit coupling $H_\mathrm{SO}=\xi\hat{\mathbf{l}}\hat{\mathbf{S}}$ to $H_{cf}$,
$\mathbf{n}$ is the orientation of the order parameter $\mathbf{L}$, and the total energy is found as the sum of the lowest three eigenvalues.
We emphasize that this model neglects spin canting, and its results should be compared with $K_{col}$.

\begin{figure}[htb]
\includegraphics[width=0.5\textwidth]{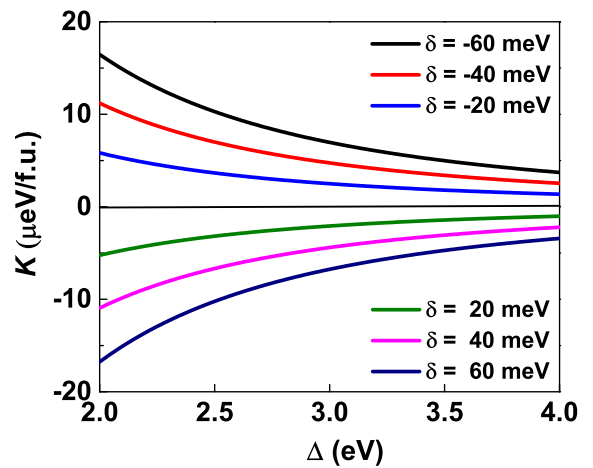}
\caption{Single electron model: MCA as a function of $\Delta$, at $\xi=35$ meV, for several values of $\delta$. The Hund exchange parameter is set to 5 eV.}
\label{splitting}
\end{figure}

Figure \ref{splitting} displays the dependence of $K_{col}$ on the crystal-field parameters $\delta$ and $\Delta$, at $\xi=35$ meV and $\epsilon=5$ eV. Clearly, positive or negative $\delta$ (i.e., singlet above or below the doublet) leads, respectively, to easy-plane or easy-axis anisotropy. Zero $\delta$ corresponds to cubic crystal field, for which $K_{col}=0$, and the magnitude of $K_{col}$ is proportional to $\delta$ (as long as $\delta\ll\Delta$).
On the other hand, $K_{col}$ depends inversely on $\Delta$, tending to zero at $\Delta\to\infty$.
These features are consistent with the first-principles results for the strain dependence of $K_{col}$ shown above in Fig.\ \ref{epi}: $\delta$ becomes more negative under tensile strain, and thereby $K_{col}$ increases.

\section{Effects of substitutional alloying}\label{dope}

Alloying can influence MCA through lattice deformations around the impurities, which affect the crystal-field splitting on the nearby Cr atoms. In addition, substitution of other transition-metal elements for Cr can result in a different electronic configuration on the impurity atoms, potentially leading to large contributions to MCA from spin-orbit coupling on those atoms.

Alloying can be introduced deliberately to tune MCA or other properties. For example, antiferromagnetic resonance measurements \cite{Foner} on (Cr$_{1-x}$Al$_{x}$)$_2$O$_3$ single crystals revealed an enhancement of MAE by Al substitution. The magnetic properties of (Cr$_{1-x}$Al$_{x}$)$_2$O$_3$ thin films have also been explored \cite{Binek2015}. Substitution of B for O can be used to increase the N\'eel temperature of \CHR\ \cite{Mu,Street} to facilitate room-temperature applications.
On the other hand, \CHR\ films for exchange-bias heterostructures are often grown on V$_2$O$_3$ \cite{Kosub}, TiO$_2$ \cite{Yuan}, Fe$_2$O$_3$ \cite{Banerjee,Shimomura}, or Co \cite{Nozaki} substrates, which can result in some degree of interfacial intermixing.

The effect of Al substitution on MCA is investigated in Sec. \ref{al}, and that of selected $3d$ and $4d$ transition-metal elements in Sec. \ref{3d}; the role of boron substitution is studied in Sec. \ref{boron}.

\subsection{Al substitution}\label{al}

\begin{figure}[htb]
\includegraphics[width=0.5\textwidth]{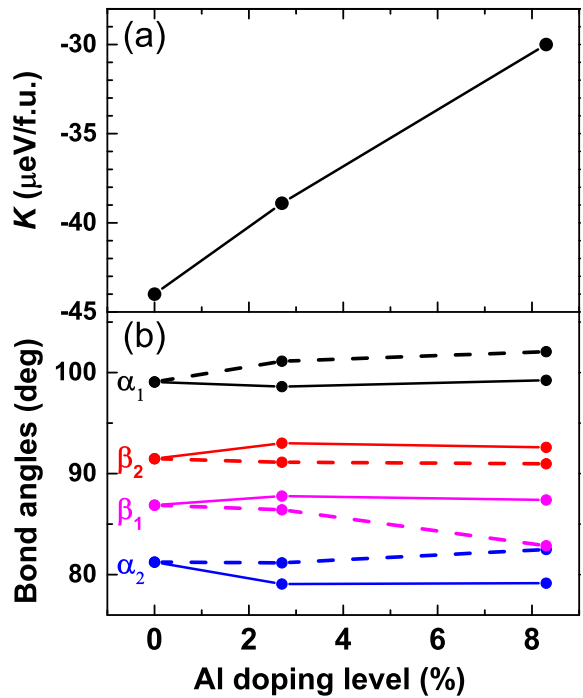}
\caption{Effect of Al doping on (a) $K$ and (b) O-Cr-O bond angles defined in Fig.\ \ref{local}. Solid and dashed lines in panel (b) correspond to Cr atoms that are, respectively, first and second-nearest neighbors of Al. For the latter, the angles are averaged over the triplets of bonds that are equivalent in pure \CHR.}
\label{alMCA}
\end{figure}

We considered substitution of one Cr atom by Al in a conventional 30-atom hexagonal supercell and in an enlarged 90-atom supercell with $\sqrt{3}\times\sqrt{3}$ translations in the $ab$ plane. This amounts to 8.3 \% and 2.8 \% Al substitution, respectively. Figure\ \ref{alMCA}a shows the calculated MCA for these two supercells, which is seen to increase at a rate of about 1.7 $\mu$eV/f.u. per 1\% of Al substitution.
Experimental measurement suggests the increase of about 0.7 $\mu$eV/f.u. per 1\% of Al substitution \cite{Foner}. Thus, the trend is predicted correctly, but the effect is overestimated.

The isovalent Al$^{3+}$ cation has no $3d$ electrons and retains the $C_{3}$ symmetry of the Al site. The impurity induces a structural deformation of the surrounding lattice. Because Al atoms are smaller than Cr, the volume of the supercell decreases by about 0.13\% per 1\% Al doping, while the $c/a$ ratio is almost unchanged (see Table \ref{bonds}). This \emph{homogeneous} strain is too weak to have an appreciable effect on MCA.

\begin{figure}
\includegraphics[width=0.5\textwidth]{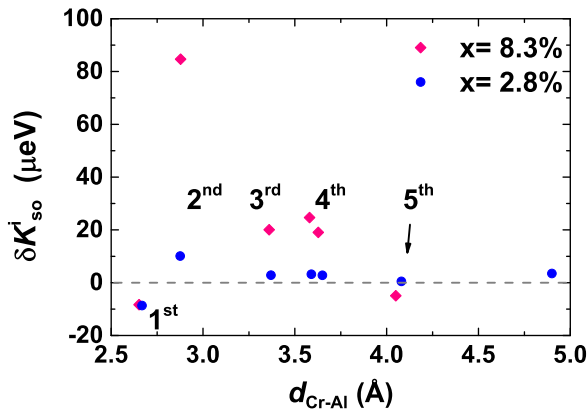}
\caption{The changes of atom-resolved anisotropies of the spin-orbit energies on one Cr site (meV) as a function of distance from Al to Cr(denoted as $d_{Cr-Al}$, from $1^{st}$ nearest shell to $5^{th}$ nearest shell), for Al doping level $x$ is 2.8\% and 8.3\%. Zero point (dashed line) corresponds to the bulk value. Note that each point is an average over Cr atoms within a certain neighboring shell.}
\label{sqrtesoc}
\end{figure}

Figure \ref{sqrtesoc} shows the site-resolved $K_\mathrm{SO}$, which was defined in Sec.\ \ref{kso}, for the Cr sites arranged in the order of increasing shortest distance from the Al impurity.
The largest contribution to the enhancement of MCA comes from the three Cr atoms that are second-nearest neighbors of Al. This enhancement is induced by the lattice deformation around the Al impurity, which effectively tunes the crystal field splitting on Cr.

The bond angles for the first and second-nearest Cr neighbors of Al are listed in Table \ref{bonds} and shown in Fig. \ref{alMCA}b. Due to the smaller size of Al compared to Cr, the O atoms shift toward it. As a result, the bond angles $\alpha_i$ increase for the second-nearest Cr (which is in the same buckled honeycomb layer with Al) and decrease for the first-nearest Cr (which is in a different layer). The changes of the $\beta_i$ angles are opposite to those of $\alpha_i$. The structural deformation is, thus, similar to tensile epitaxial strain on the second-nearest Cr atoms and compressive on the first-nearest one (see Table \ref{bonds}). The changes in $K_\mathrm{SO}$ on these atoms are thus attributable to crystal-field effects that were discussed above.

\subsection{Substitution of $3d$ and $4d$ elements for Cr} \label{3d}

In this section, we consider the effects of Cr substitution by $3d$ (Ti, V, Fe, Co, Ni) and $4d$ (Nb, Mo, W) elements, replacing one Cr atom by an impurity in the 30-atom hexagonal cell.
In order to obtain an accurate electron structure and MCA for doped \CHR, we used the \emph{ab-initio} linear response method \cite{Cococcioni} to evaluate the $U_\mathrm{eff}$ parameters for the $3d$ impurities. For the $4d$ elements the $U_\mathrm{eff}$ parameters were obtained using the constrained occupation approach within the full-potential linear augmented plane-wave method \cite{Blugel,Shi}.
These values of $U_\mathrm{eff}$ are listed in Table \ref{cation}.

The results of MCA calculations are listed in Table \ref{cation}. The most striking effect on MCA comes from V and Nb doping: $K$ is reduced at a rate of 216 or 156 $\mu$eV/f.u. per 1\% substitution of V or Nb, respectively. This strong effect is not attributable to lattice distortion, which we explicitly rule out by recalculating the MCA of pure \CHR\ in the distorted geometry for the given substitution (denoted as $K^{\dagger}$ in Table \ref{cation}). We see that $K^{\dagger}$ for V and Nb-substituted structures is close to the value of $K$ for pure \CHR. Further, the large negative MCA in V and Nb-substituted systems is almost entirely due to the large negative $K_\mathrm{SO}$ on the impurity atom, and it occurs in the collinear approximation as well (see the $K_{col}$ values in Table \ref{cation}).
Therefore, to understand the strong effect of V and Nb on MCA, we focus on $K_{col}$.

\begin{table*}[htb]
\caption {Properties of Cr$_2$O$_3$ with a substitution of one Cr atom by a dopant $X$ in a 30-atom supercell (8.3\% substitution). $d_1$, $d_2$: $X$-O bond lengths (\AA); $\alpha_{1,2}$, $\beta_{1,2}$: O-$X$-O bond angles (deg) shown in Fig.\ \ref{local}; $U$, $J$, $U_\mathrm{eff}$: effective interaction parameters (eV) for the atom $X$; $K$ ($\mu$eV/f.u.): MCA with full spin relaxation; $K_{col}$ ($\mu$eV/f.u.): MCA obtained assuming collinear spin states; $K^\dagger$ and $K^\dagger_{col}$ ($\mu$eV/f.u.): same as $K$ and $K_{col}$ but with $X$ replaced back by Cr with the structure optimized in the presence of $X$; $K_\mathrm{SO}^{X}$ ($\mu$eV): see Sec.\ \ref{kso}; $\mu_{X}$ ($\mu_B$): local magnetic moment on the defect atom $X$; $\eta$ ($\mu$eV/f.u.): rate of change of $K$ per 1\% substitution of Cr by $X$.
Bond lengths and angles are not listed for Jahn-Teller elements Ti, Ni, and Zr.}
\begin{tabular}{|c|c|c|c|c|c|c|c|c|c|}
\hline
$X$      & Cr & Ti &V &Fe &Co& Ni & Zr & Nb & Mo  \\
\hline
$d_1$     		    &  1.96      &  --        &  1.976      & 1.937     & 1.898    & --        &   --         & 	2.058    & 	2.100   \\
$d_2$  		    &   2.01     &  --        &   2.046     &	2.061     &1.918     & --        &      --	  &  2.126 	    &  2.062 \\
\hline
$\alpha_1$	   & 99.07      & --         &  100.2      & 101.47  &98.21      & --        &  -- 	  & 99.62       & 	98.48 \\
$\alpha_2$	   &  81.23     &  --        &   79.68     & 78.97    & 82.97     & --        & --	           & 77.95      & 	80.53  \\
\hline
$\beta_1$      	   & 86.87     &  --         & 86.88	     & 86.44    &86.78     &  --        &   --         &	88.13   & 87.42 \\
$\beta_2$ 	   &  91.47    &   --        & 91.31	     & 90.89    &91.07     & --         &  --	  &  92.36	    & 92.27\\ 	
\hline
$c$           		   &  13.553  & 13.609 &  13.580    &	13.555  & 13.489 & 13.502 & 13.711   & 	13.747 &	13.669 \\
$a$        		   &  4.945    & 4.949   &    4.949    & 	4.944    & 4.920   &  4.933  &  4.997    &	4.964   & 4.963\\
$c/a$        		   &  2.741   & 2.750   &    2.744    &2.742     &2.742    & 2.737   & 2.744	  & 2.770	   & 2.754 \\
\hline
$U$                	   &4.0          & 3.95     &  4.24        &  5.35     & 5.89     & 6.54    & $1.73 $  & $2.21$  & $3.16$    \\
$J$                      &0.58        & 0.47     & 0.54         & 0.75      & 0.81     & 0.85    &  $0.70$  & $0.55$  & $0.53$     \\
$U_\mathrm{eff}$& 3.4         & 3.5      & 3.7            & 4.6        & 5.1       & 5.7      & 1.0         &  1.7      & 2.6 \\
\hline
$\mu_X$         &  2.91    & 0.82     &    1.92      &4.24       &0.04     & 1.07     & 0.04	  & 1.20	   & 2.40 \\
\hline
$K^\dagger$ ($K^\dagger_{col}$)       & --   &$-42$ ($-1.4$) &  $-47$ (1.7) & $-38$ (8)   &  $-35$ (1.7) & $-41$ (4.4)  & $-33$ (9)   &  $-43$ ($-20$)    & $-64$ ($-10$)   \\
$K$ ($K_{col}$)   &$-44$ (2.5)          &$-98$ ($-79$)&$-1820$ ($-1240$) & $-53$ ($-1.7$) &$-15$ (10)  &  $-56$ ($-13$)  & 2 (15)   & $-1360$ ($-1350$) & $-66$ ($-50$)  \\
$K_{SO}^X$         &  $-24$    & $-380$     &    $-5560$      &$-85$       &$-48$     & $-63$     & $-3$	  & $-5980$	   & $-306$ \\
$\eta$         &  0    & $-6.5$     &    $-213$      &$-1.1$       &3.5     & $-1.4$     & 5.5	  & $-159$	   & $-2.6$ \\
\hline
\end{tabular}
\label{cation}
\end{table*}

\begin{figure}[htb]
\includegraphics[width=0.5\textwidth]{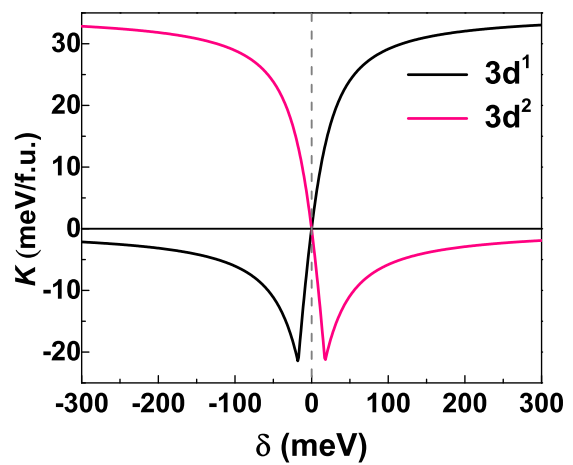}
\caption{MCA as a function of the $t_{2g}$ crystal-field parameter $\delta$, at $\xi=35$ meV, for $3d^1$ and $3d^2$ configurations. Note the different scale (meV) for MCA compared to Fig.\ \ref{splitting}.}
\label{filling}
\end{figure}

Both V$^{3+}$ and Nb$^{3+}$ are $3d^2$ ions. Repeating the calculations of MCA for the simple model described in Sec. \ref{strain}, but with either 1 or 2 electrons in the $3d$ shell instead of 3, we find a very large MCA that strongly depends on the $t_{2g}$ splitting parameter $\delta$ (see Fig. \ref{filling}).

The mechanism of MCA can be understood as follows. The singlet has character $m=0$ and the doublet $m=\pm 1,\pm 2$ (in the hexagonal frame). For $\mathbf{L}\parallel \hat z$, the spin-orbit coupling linearly splits the doublet without affecting the singlet. The upper branch of the split doublet crosses the singlet state at $\xi=2\delta$. For $\mathbf{L}\perp \hat z$, the singlet is mixed with the doublet, also splitting the latter, and there are no level crossings.

For the $3d^2$ configuration, at $\delta>2\xi$ the spin-orbit coupling only reduces the energy of the in-plane spin configuration. This reduction is inversely proportional to $\delta$ and corresponds to the behavior of $K$ at large $\delta$ in Fig.\ \ref{filling}. Once $\delta$ is reduced down to $2\xi$, the level crossing leads to an abrupt switch to a linear region in $K(\delta)$. At $\delta=0$ the crystal field is cubic and, therefore, $K$ vanishes. At negative $\delta$, $K$ continues to rise and approaches a large positive value at $|\delta|\gg\xi$. In this limit, only the splitting of the doublet at $\mathbf{L}\parallel \hat z$ is important. Since this doublet is half-filled, there is a large energy gain for $\mathbf{L}\parallel \hat z$, leading to a large easy-axis anisotropy.

The situation for the $3d^1$ configuration at splitting $\delta$ is almost identical to $3d^2$ at $-\delta$; the tiny difference, which is unnoticeable in Fig.\ \ref{filling}, comes only from the mixing with higher-lying states.

This fully localized ionic model neglects hybridization, which should considerably reduce the anisotropy and smear out the cusps at $|\delta|=2\xi$. However, strong negative MCA for the $3d^2$ impurities V and Nb is consistent with the prediction of this model for $\delta>0$.

Further, the $3d^2$ configuration at $\delta<0$ and the $3d^1$ configuration at $\delta>0$ are degenerate in the absence of spin-orbit coupling and are, therefore, expected to undergo Jahn-Teller distortion. In contrast, $3d^2$ at $\delta>0$ and $3d^1$ at $\delta<0$ are not subject to the Jahn-Teller effect. The fact that $3d^2$ impurities V and Nb in \CHR\ don't, while $3d^1$ impurities Ti and Zr do, experience Jahn-Teller distortion \cite{Mu} is consistent with $\delta$ being positive for all these dopants and with the finding of large and negative $K$ for V and Nb. On the other hand, the distortion strongly changes the crystal field for Ti and Zr, invalidating the prediction of large positive MCA for the $3d^1$ configuration (Fig.\ \ref{filling}). As seen in Table \ref{cation}, Zr (Ti) increases (reduces) $K$ at a rate of 5.9 $\mu$eV/f.u. (6.5 $\mu$eV/f.u.) per 1\% substitution.
For both Ti and Zr, the value of $K^\dagger$, which is obtained by replacing the impurity back by Cr while maintaining the deformed structure, is similar to ideal \CHR. This indicates that the effect of Ti and Zr on MCA comes from the impurity atom itself.

Consistent with the above analysis, V and Nb atoms carry large orbital moments for the in-plane spin configuration ($-0.78$ and $-0.33\ \mu_B$, respectively), but not for out-of-plane configuration ($-0.04$ and $-0.06\ \mu_B$). (The negative signs indicate that orbital and spin moments are antiparallel.)
Further, if one electron is added to the V-doped \CHR\ supercell by introducing a homogeneous positive background charge, V$^{3+}$ turns into V$^{2+}$ with a $3d^3$ configuration. This system has an MCA of $-3.3$ $\mu$eV/f.u., which is larger than bulk \CHR. This confirms that the large negative MCA results only from V$^{3+}$ and Nb$^{3+}$ ions in the $3d^2$ electronic configuration.

As seen in Table \ref{cation}, $K$ and $K_{col}$ are almost the same for Nb, while there is a large difference for V, which comes from the spin canting in the in-plane spin configuration. The canting angle for Nb is only 0.4$^{\circ}$, while for V it is as large as 2.2$^{\circ}$.

In addition to Ti and Zr, the Ni impurity with the low-spin $3d^7$ configuration also experiences a Jahn-Teller distortion, which originates in the single occupation of the degenerate $e_g$ orbitals \cite{Mu}, while the $t_{2g}$ states are filled for both spins channels. Ni substitution has a very small negative effect on MCA.

\begin{figure}
\includegraphics[width=0.48\textwidth]{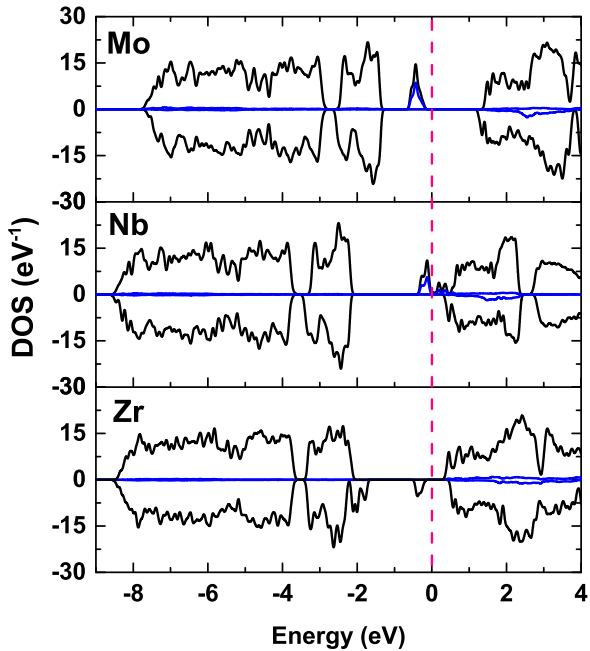}
\caption{Total density of states (DOS) of Zr, Nb and Mo doped \CHR, shown in black lines. The 4$d$ impurity states are shown in blue lines. $\pm$ sign of DOS indicates different spin channels. The Fermi energy has been shifted to 0 eV. }
\label{4ddos}
\end{figure}

For isovalent Mo substitution, the reduction of $K^\dagger_{col}$ relative to pure \CHR\ (from 2.5 to $-10$ $\mu$eV/f.u.) reflects the local strain effect, which primarily affects the second-nearest Cr neighbors of Mo (as follows from the analysis of $K_\mathrm{SO}$). The substitution of Mo in the distorted structure further reduces MCA to $-50$ $\mu$eV/f.u.\ thanks to a large negative $K_\mathrm{SO}$ for the Mo atom. The contribution of spin canting to $K$ is reduced in Mo-doped \CHR\ due to the enhancement of the exchange interaction near the Mo atom. Overall, Mo reduces MCA at a small rate of 2.6 $\mu$eV/f.u. per 1\% Mo substitution.

The exchange interaction in the Mo-doped case is noteworthy. The effect of alloying on the N\'eel temperature can be estimated from the values of the exchange energies $E_i$, which represent the energy cost to reverse the local moment on atom $i$ \cite{Mu}. In pure Cr$_2$O$_3$ this energy is 120 meV \cite{Mu}. It is increased to 620 meV for the Mo dopant, 240 meV for its nearest Cr neighbors, and 420 meV for the second Cr neighbors of Mo. This enhancement can be attributed to the impurity state in the band gap (see Fig. \ref{4ddos}), which mediates the exchange interaction.
A rough estimate based on the mean-field approximation suggests that 1\% Mo substitution increases the N\'eel temperature by about 8\%, which is comparable to the effect of B substitution for O \cite{Mu,Street}.

The isovalent Co$^{3+}$ impurity ($3d^6$ configuration) prefers the low-spin state with fully filled $t_{2g}$ states and empty $e_g$ states \cite{Mu}. The spin moment on Co is only 0.04 $\mu_B$. A somewhat enhanced $K^\dagger$ indicates that the pure structural relaxation effect increases the MCA, but an even larger enhancement comes from $K_\mathrm{SO}$ on Co. Overall, Co increases MCA by 3.6 $\mu$eV/f.u. per 1\% substitution. The enhancement of MCA under Co doping may explain the observation \cite{Shiratsuchi} of an enhanced perpendicular anisotropy in Co/Cr$_2$O$_3$ bilayers.

Substitution of Cr by Fe slightly reduces $K$, which could be the origin of the reduced coercivity field in Cr$_2$O$_3$/Fe$_2$O$_3$ compared to Cr$_2$O$_3$/Pt \cite{Shimomura}.

For most practical applications, \CHR\ alloyed with other elements should retain its insulating properties. Impurities of $3d$ transition metals introduce impurity levels inside the band gap \cite{Mu}, which are likely to degrade the insulating properties for sizable levels of substitution. Therefore, transition-metal substitutions are likely to be useful only if they have a strong effect on MCA.

Our results suggests that MCA of \CHR\ can be effectively reduced, or even switched to easy-plane, by a very small substitution of V or Nb on the order of 0.05\%. The density of states for Nb-doped \CHR, as seen in Fig. \ref{4ddos}, shows filled impurity states close to the bottom of the conduction band, which may act as electron donors. In contrast, the excitation gap in V-doped \CHR\ is almost 2 eV wide, with filled impurity states close to the top of the valence band and empty impurity states close to the bottom of the conduction band \cite{Mu}. This difference is due to a much larger on-site Coulomb interaction in the $3d$ shell of V. The lack of easily excitable impurity states in V-doped \CHR\ makes V preferable to Nb as an alloying element if reduced MCA is desired. For \CHR\ films grown on a V$_2$O$_3$ substrate \cite{Kosub}, even a small Cr/V interdiffusion can strongly reduce MCA or turn it to easy-plane.

The strongest positive effect on MCA, among transition elements, is from Zr substitution, which is about 3 times more effective than Al. However, similar to Nb, Zr has shallow electron donor states (see Fig. \ref{4ddos}), which are likely to degrade the insulating properties. Cobalt may be a preferable alternative: its effect on MCA is predicted to be twice stronger than Al, while the insulating properties may be expected to be preserved thanks to the wide excitation gap for the Co$^{3+}$ ions in \CHR\ \cite{Mu}.

\subsection{Boron substitution for oxygen}\label{boron}

Substitution of boron on the oxygen sublattice in \CHR, on the level of a few percent, was shown to considerably increase the N\'eel temperature \cite{Mu,Street}. Given the potential utility of this material for applications, it is important to examine the effect of boron substitution on the magnetic anisotropy. We focus on the B$^{2-}$ and B$^{1-}$ charge states, which, according to theoretical calculations \cite{Mu}, are expected to enhance the N\'eel temperature.

The neutral B$^{2-}$ impurity strongly distorts the local geometry \cite{Mu}. As seen from Table \ref{bMCA}, these impurities decrease MCA. The analysis of $K_\mathrm{SO}$ indicates this decrease is primarily due to the modified crystal field on the four Cr atoms that are bonded with B. Comparison of $K$ and $K^\dagger$ shows that the states of B somewhat increase MCA relative to O, but this effect only partially compensates the reduction of MCA due to the structural distortion.

\begin{table}[htb]
\caption {$K$ and related quantities ($\mu$eV/f.u.; see Table \ref{cation} for definitions) for \CHR\ with 5.6\% substitution of B for O. The spin structure for B$^{3-}$ could not be converged.}
\begin{tabular}{|c|c|c|c|c|}
\hline
 MCA     & bulk & B$^{2-}$ & B$^{1-}$  & B$^{3-}$   \\
\hline
$K$ ($K_{col}$)         & $-44$ (2.5) & $-62$ ($-2.2$)& 26 (60) &   --- ($-206$) \\
$K^\dagger$ ($K^\dagger_{col}$) & $-44$ (2.5)& $-71$ ($-12$) &  $-30$ (14) & $-80$ ($-21$)   \\
\hline
\end{tabular}
\label{bMCA}
\end{table}

The B$^{1-}$ charged state was enforced by introducing a homogeneous background charge to the supercell. It is seen from Table \ref{bMCA} that B$^{1-}$ impurities increase MCA at a rate of $+8.4$ $\mu$eV/f.u. per 1\% substitution. This increase is almost entirely due to the presence of B electronic states, because $K^\dagger$, which accounts for the structural distortion alone, is only slightly increased compared to pure \CHR. It originates from the four nearby Cr sites, as follows from the analysis of $K_\mathrm{SO}$.

\section{Magnetic dipolar interactions}

A calculation \cite{Artman} based on the experimental structure for \CHR\ estimated the contribution from the magnetic dipolar interaction to MAE at $K_{dd} = 3 $ $\mu$eV/f.u. This value is comparable to and of the same sign as the experimental MCA. Here we study the effect of epitaxial strain on $K_{dd}$, using the theoretically optimized structures.

The interaction energy of two localized magnetic dipoles $\boldsymbol{\mu}_i$ and $\boldsymbol{\mu}_j$ separated by $\mathbf{r}_{ij}=\mathbf{r}_j-\mathbf{r}_i$ is
\begin{equation}
E_{ij}=-\frac{\mu_0}{4\pi} \frac{3(\boldsymbol{\mu}_i \cdot\mathbf{n}_{ij})(\boldsymbol{\mu}_j\cdot \mathbf{n}_{ij})-\boldsymbol{\mu}_i\cdot \boldsymbol{\mu}_j}{r_{ij}^3},
\end{equation}
where $\mathbf{n}_{ij}=\mathbf{r}_{ij}/r_{ij}$.
The sum over all pairs of dipoles gives the total magnetostatic energy. In the calculations for \CHR\ we assumed $\mu_i=3$ $\mu_B$.

For the unstrained theoretical structure we found $K_{dd}=2.4$ $\mu$eV/f.u., in good agreement with Ref.\ \onlinecite{Artman}. The results as a function of epitaxial strain are shown in Fig.\ \ref{epi}. We see that $K_{dd}$ is only comparable to MCA in the unstressed system, and it has an opposite trend under strain.Thus, it does not materially affect the strain dependence of MAE.

\section{Effect of electric field} \label{E_field}

\begin{figure}[htb]
\includegraphics[width=0.5\textwidth]{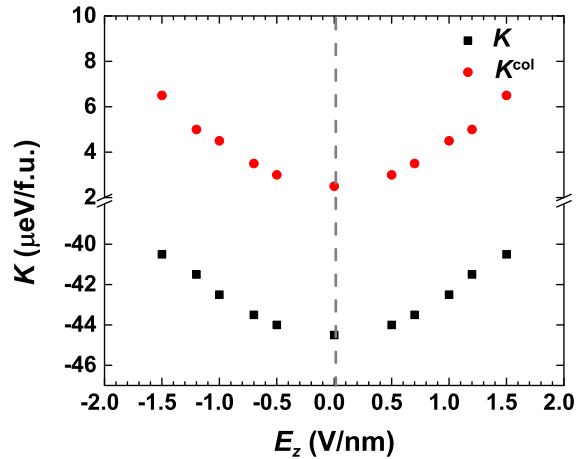}
\caption{Dependence of $K$ and $K^{col}$ on the electric field applied along the rhombohedral axis.}
\label{MCA_E_ion}
\end{figure}

Here we investigate the response of MCA to a uniform electric field applied along the rhombohedral axis. Figure \ref{MCA_E_ion} shows $K$ and $K^{col}$ in the presence of such field. The dependence is quadratic, and the response coefficient is similar for $K$ and $K^{col}$: $K=-44.4+1.8E_z^2$ and $K_{col}=2.5+1.7E^2$ (in units of $\mu$eV/f.u., with $E_z$ in V/nm). The dependence is rather weak thanks to the rigidity of the crystal lattice. For example, a 1 V/nm electric field changes the Cr-O bond length by only about 1\%. We found that the effect of the electric field on MCA is entirely due to the induced ionic displacements; the electronic contribution is negligible.

Although the effect of electric field on MCA is relatively weak, it could be used to switch the anisotropy from easy-plane to easy-axis by voltage in \CHR\ that is doped (for example, by vanadium) slightly beyond the point where MCA vanishes. Such voltage control of magnetic anisotropy in an antiferromagnet could lead to interesting applications.

\section{Conclusions}\label{concl}

The influence of epitaxial strain, substitutional alloying, and electric field on the MAE of \CHR\ has been studied using the DFT+$U$ method. The MAE of pure, unstrained \CHR\ is very small, because the crystal field of the Cr ion is close to cubic, and the calculations fail to reproduce the correct positive sign of MCA. Tensile epitaxial strain increases MCA at a rate of about 60 $\mu$eV/f.u. per 1\% strain, which is due to the changing crystal-field splitting of the $t_{2g}$ states. The large response of MCA makes strain an effective tuning parameter for applications.

Substitution of Cr by Al is predicted to increase MCA at the rate of 1.7 $\mu$eV/f.u. per 1\% substitution, in qualitative agreement with experiment. This effect is mediated by the structural distortion around the Al impurities. Among the $3d$ and $4d$ transition-metal alloying elements, V and Nb have the strongest effect on MCA, reducing $K$ at very large rates of about 220 and 160 $\mu$eV/f.u. per 1\% substitution, respectively. This large effect is due to the partial ($3d^2$) filling of the $t_{2g}$ states, which enables strong in-plane anisotropy thanks to a spin-orbital selection rule. On the other hand, Co and Zr are predicted to increase MCA at a moderate rate that is 2-3 times larger than that of Al. For applications where the alloyed \CHR\ needs to retain good insulating properties, V is the element of choice for reducing MAE or switching it to easy-plane, while Al or Co are preferred for increasing MAE.

Boron substitution for O increases the N\'eel temperature of \CHR\ and is, therefore, desirable for applications. We predict that neutral B$^{2-}$ impurities tend to moderately decrease MAE, while B$^{1-}$ increase it considerably.

Magnetostatic contribution to MAE is estimated at $2.4$ $\mu$eV/f.u.\ in unstrained \CHR, which is comparable to the experimental MAE of 6 $\mu$eV/f.u. However, the magnetostatic contribution remains small under epitaxial strain, while the magnetocrystalline anisotropy is strongly modified by it.

The electric field applied along the hexagonal axis is predicted to increase MCA quadratically ($K\sim E_z^2$). The field $E_z\sim 1$ V/nm increases MAE by about 1.8 $\mu$eV/f.u. If \CHR\ is alloyed with V slightly beyond the point where MAE vanishes, the electric field could be used to switch MAE from easy-plane to easy-axis. \\

\begin{acknowledgments}
\label{thanks}
We are grateful to Christian Binek, James Glasbrenner and Satoshi Okamoto for useful discussions and to Martijn Marsman for assistance with the VASP code. This work was supported by the National Science Foundation (NSF) through the Nebraska Materials Research Science and Engineering Center (MRSEC) (Grant No.\ DMR-1420645), and by the Nanoelectronics Research Corporation (NERC), a wholly-owned subsidiary of the Semiconductor Research Corporation (SRC), through the Center for Nanoferroic Devices (CNFD), a SRC-NRI Nanoelectronics Research Initiative Center (Task ID 2398.001). Calculations were performed utilizing the Holland Computing Center of the University of Nebraska, which receives support from the Nebraska Research Initiative.
\end{acknowledgments}

\end{document}